\documentclass[prd,twocolumn,superscriptaddress,altaffilletter,amssymb,showpacs,nofootinbib]{revtex4}
\usepackage[dvips]{graphicx}
\usepackage{amsmath}
\newcommand{\be}{\begin{equation}}
\newcommand{\ee}{\end{equation}}
\newcommand{\bea}{\begin{eqnarray}}
\newcommand{\eea}{\end{eqnarray}}
\newcommand{\vphi}{\varphi}

\usepackage{amssymb}
\usepackage[dvips]{graphicx}
\usepackage{amsmath}
\usepackage{amssymb}
\usepackage{times}
\usepackage[T1]{fontenc}
\usepackage{tikz}
\usepackage{latexsym}
\usepackage{epsfig}
\usepackage{amsmath}
\usepackage{amsfonts}
\usepackage{graphicx}
\usepackage{makeidx}
\usepackage[spanish, activeacute]{babel}
\usepackage[latin1]{inputenc}

\begin{document}

\title{Vacuum Structure of Cosmologically Viable Quadratic Modifications of Gravity\\that are Functions of the Gauss-Bonnet Invariant}

\author{Israel Quiros}\email{iquiros@fisica.ugto.mx}\affiliation{Divisi\'on de Ciencias e Ingenier\'ia de la Universidad de Guanajuato, A.P. 150, 37150, Le\'on, Guanajuato, M\'exico.}

\author{Eduardo Test\'e}\email{eteste@fises.fisica.uh.cu}\affiliation{Departamento de F\'{\i}sica, Universidad de La Habana, Habana, Cuba.}

\date{\today}

\begin{abstract}
We perform a thorough study of the theoretical consistency of recently proposed, viable, quadratic modifications of gravity that are functions of the the Gauss-Bonnet invariant, regarding the stability of their perturbations around vacuum, maximally symmetric spaces of constant curvature. We pay special attention, in particular, to the investigation of pathological instabilities associated with the occurrence of propagating spin-0 tachyon modes, and with the development of a graviton ghost. The latter effect is associated with the known ''Ricci stability'' issue, well studied in $f(R)$-theories. Within quadratic modifications of gravity it is discussed for the first time. Special attention is paid to the requirement of non-negativity of the effective gravitational coupling, which warrants that the graviton is not a negative-norm state. It is demonstrated that, several theories that pass the cosmological as well as the solar system tests, have to be rule out on the basis of the unavoidable character of these pathological instabilities.  
\end{abstract}


\maketitle

\section{Introduction}\label{introduction}

Attempts to modify the Einstein-Hilbert (EH) action of general relativity (GR) have been motivated by a number of reasons. In particular, renormalization at one-loop demands that the Einstein-Hilbert action be supplemented by higher order curvature terms \cite{udw}. Besides, when quantum corrections or string theory are taken into account, the effective low energy action for pure gravity admits higher order curvature invariants \cite{qstring}. More recently it has been suggested that the present cosmic speed-up could have its origin in -- among other possibilities -- corrections to the GR equations of motion, generated by non-linear contributions of the scalar curvature $R$ in the pure gravity Lagrangian of $f(R)$ theories \cite{odintsov1,frspeedup,positive,carroll}. Next in degree of complexity are the so called $F(X,Y,Z)$-theories \cite{carroll,stelle,ovrut,solganik,navarro,akoleyen,chiba}, where, for simplicity of writing, we have introduced the field variables $X\equiv R$- the curvature scalar, $Y\equiv R_{\mu\nu}R^{\mu\nu}$, and $Z\equiv R_{\mu\nu\sigma\lambda}R^{\mu\nu\sigma\lambda}$. The gravitational spectrum of the linearization of these theories consists of a massless spin-2 graviton plus two additional gravitational propagating degrees of freedom: a massive spin-0 excitation, and a massive spin-2 propagating mode. The latter appears to be a ghost mode associated with the Weyl curvature invariant $C^2\equiv C_{\mu\nu\sigma\lambda}C^{\mu\nu\sigma\lambda}$ \cite{stelle,ovrut,solganik}. Notwithstanding, there are ways to overcome (or at least to smooth out) the consequences of the would be massive spin-2 ghost mode as, for instance, in case $F(X,Y,Z)=F(X,{\cal G})$, where ${\cal G}=X^2-4Y+Z$ is the Gauss-Bonnet (GB) invariant. 

It has been demonstrated that quadratic modifications of gravity where the GB invariant enters as a function $f({\cal G})$ added to the gravitational action: $F(X,{\cal G})=X+f({\cal G})$, might represent a candidate for effective dark-energy \cite{nojiri,odintsov1,odintsov2}. Examples of such quadratic modifications are those where powers of the GB invariant are involved. In Re. \cite{odintsov2}, for instance, models of the kind $f({\cal G})=f_0|{\cal G}|^\beta$ were considered. In that reference, the authors showed that, when arbitrary functions of the GB-invariant are added to the GR action, the resulting theories are endowed with a quite rich cosmological structure: they may naturally lead to an effective cosmological constant, quintessence or phantom cosmic acceleration, with a possibility for the transition from deceleration to acceleration. It has been also demonstrated that these theories are viable, since they comply with the Solar System constraints. Specific properties of $f({\cal G})$ gravity in a de Sitter universe, such as dS and SdS solutions, their entropy and its explicit one-loop quantization were also studied in \cite{odintsov2}. Additionally, the issue of a possible solution of the hierarchy problem in modified gravities was addressed too. A related, yet simpler, class of models is given by the following action:\footnote{This is a particular class of models $f({\cal G})=\mu |{\cal G}|^n$ \cite{odintsov2}, when the GB-invariant is always positive.}

\be S=\int\frac{d^4x\sqrt{|g|}}{16\pi G_N}(X+\mu{\cal G}^n),\label{powers}\ee where $G_N$ is the Newton's constant and the parameter $\mu$ can be either positive or negative. For positive values of the parameter $n$, the above model can be consistent with solar-system tests for $n\lesssim 0.074$, if the GB term is to be the cause of the present speedup of the cosmic expansion \cite{davis}. For $n<0$ the model (\ref{powers}) is not cosmologically viable due to the existence of separatrices between the radiation-dominated, and dark energy-dominated stages of the cosmic evolution \cite{de felice1}.

Recently several new $F(X,{\cal G})$-models have been proposed \cite{dft,zcs} that are peculiar functions of the GB invariant, such as $\arctan$, $\exp$, etc. In \cite{dft}, three particular models have been proposed ($F(X,{\cal G})=X+f({\cal G})$):

\bea &&f({\cal G})=\frac{\lambda}{\sqrt\beta}{\cal G}\arctan\left(\frac{{\cal G}}{\beta}\right)\nonumber\\
&&\;\;\;\;\;\;\;\;\;\;\;\;\;\;-\frac{\lambda\sqrt\beta}{2}\ln\left(1+\frac{{\cal G}^2}{\beta^2}\right)-\alpha\lambda\sqrt\beta,\label{dft A}\\
&&f({\cal G})=\frac{\lambda}{\sqrt\beta}{\cal G}\arctan\left(\frac{{\cal G}}{\beta}\right)-\alpha\lambda\sqrt\beta,\label{dft B}\\
&&f({\cal G})=\lambda\sqrt\beta\ln\left[\cosh\left(\frac{{\cal G}}{\beta}\right)\right]-\alpha\lambda\sqrt\beta,\label{dft C}\eea where $\lambda$, $\alpha$, and $\beta$ are positive constants. Meanwhile, in Ref. \cite{zcs} the following models have been explored:

\bea &&f({\cal G})=\alpha\sqrt{\cal G}+\beta\sqrt[4]{\cal G},\label{zcs A}\\
&&f({\cal G})=\alpha\left({\cal G}^{3/4}-\beta\right)^{2/3},\label{zcs B}\\
&&f({\cal G})=\alpha\sqrt{\cal G}\exp\left(\frac{\beta}{\cal G}\right),\label{zcs C}\eea where, as before, $\alpha$ and $\beta$ are constants. The above models have been shown to be cosmologically viable and compatible with solar system tests. In Ref. \cite{cosmo tests}, for instance, the models (\ref{dft A}-\ref{zcs C}) were compared to combined data-sets of supernovae, baryon acoustic oscillations, and constraints from the CMB surface of last scattering. The authors found that these models provide data-fits that are close to those of the LCDM concordance model. It is suggested that higher order gravity models that pass the mentioned tests, need to be compared to constraints from large scale structure and full CMB analysis \cite{cosmo tests}.

A similar, seemingly positive situation, arises with theories of the kind (\ref{powers}). Actually, a generalization of the theory depicted by the action (\ref{powers}) for negative $\mu=-\bar\mu^{2-4n}$ and $n<0$, which is singular in the limit of vanishing curvature, has been discussed in details in references \cite{navarro,akoleyen} (see also \cite{chiba}). It has been demonstrated, in particular, that although gravity is modified at large distances, one recovers an acceptable Newtonian limit at small distances \cite{akoleyen}. It has been discussed there that, with the values of the parameter $\bar\mu$ necessary to explain the cosmic acceleration ($\bar\mu\sim H_0$), the correction to the Newtonian potential is negligible for solar system scales but becomes important at galactic lengths. In consequence, although having a negligible effect on the solar system, this kind of modifications would have implications for the dark matter problem \cite{navarro,akoleyen}. The above behavior can be explained as follows. Even if one chooses a theory (\ref{powers}), which is spin-2, ghost-free (it corresponds to the choice $a=c$, $b=-4c$ in the model of Ref. \cite{navarro,akoleyen}), there is an additional spin-0 degree of freedom in the resulting theory. It then arises that the mass of this extra excitation depends on the background in such a way that it effectively decouples when close to any matter source (the Sun, galaxies, etc.), but in vacuum it is of the order of the Hubble scale. However, a careful study of the vacuum structure of the corresponding linearized theory shows that the above long-distance modification of gravity\footnote{Recall that we are considering the particular case given by (\ref{powers}) with both $\mu<0$ and $n<0$ negative quantities.} develops a ghost graviton, and also, a tachyon instability associated with negative mass squared of the spin-0 propagating perturbation \cite{quiros-urena}.\footnote{Here we have in mind perturbations around vacuum maximally symmetric spaces of constant curvature.} Hence this theory has to be rule out as inconsistent. This demonstrates the need for a careful investigation of the possible pathological instabilities that can be associated with the linearization of the above quadratic modifications of gravity (\ref{powers}-\ref{zcs C}) around vacuum, maximally symmetric spaces of constant curvature.

Aim of the present paper is, precisely, to carefully investigate the vacuum structure of the proposed $F(X,{\cal G})$-theories by expanding the corresponding Lagrangian densities about vacuum, maximally symmetric spaces of constant curvature. Then we look at the spectrum of perturbations of the metric by comparing with known results \cite{stelle,ovrut,solganik}. We pay special attention to checking the models regarding absence of Ricci instabilities, as well as spin-0 tachyon instabilities, and, also, to checking non-negativity of the effective gravitational coupling, an implicit requirement that is not always carefully checked. Ricci instability is usually associated with the development of a ghost graviton and, although it has been well established for $f(X)$-theories, within $F(X,Y,Z)$ gravity theories, as long as we know, this issue has not been discussed before.

\section{Vacuum Instabilities}\label{vacuum stability}

Stability issues are central in the study of higher order modifications of general relativity, since these are plagued by several kinds of instabilities, some of which are catastrophic, leading to subsequent ruling out of the corresponding theories. Amongst these is the fundamental Ostrogradski instability, based on the powerful no-go theorem of the same name \cite{woodard}: ``there is a linear instability in the Hamiltonians associated with Lagrangians which depend upon more than one time derivative in such a way that the dependence cannot be eliminated by partial integration''. This result is general and can be extended to higher order derivatives in general. As a consequence, the only Ostrogradski-stable higher order modifications of Einstein-Hilbert action are those in the form of an $f(X)$ function \cite{woodard}. This result alone might rule out any intent to consider quadratic modifications such as the ones of interest here. However, the subject is subtle and, in the last instance, consideration of such theories can shed more light on the stability issue. 

In order to discuss such subtle issues, usually, one expands the action of the theory in the neighborhood of background (vacuum) spaces of constant curvature. If one considers general actions of the kind 

\be S=\int\frac{d^4x\sqrt{|g|}}{16\pi G_N}F(X,Y,Z),\label{action}\ee then, Taylor expanding the function $F(X,Y,Z)$ around maximally symmetric spaces of constant curvature $X=X_0$, $Y=Y_0=X_0^2/4$, $Z=Z_0=X_0^2/6$ (the subscript "$0$" means the given invariant is evaluated at the vacuum, constant curvature background), and keeping terms up to ${\cal O}(3)$, one gets (see \cite{quiros-urena} for details): 

\be F(X,Y,Z)=\lambda+\alpha X+\frac{\beta}{6}X^2+\frac{\gamma}{2} C^2\;,\label{taylor}\ee where $$C^2\equiv C_{\mu\nu\sigma\lambda}C^{\mu\nu\sigma\lambda}=Z-2Y+\frac{1}{3}X^2$$ is the Weyl curvature invariant, and the coefficients are

\bea &&\lambda\equiv F_0-X_0 F_X^0-Y_0 F_Y^0-Z_0 F_Z^0+\frac{X_0^2}{2}F_{XX}^0\;,\nonumber\\
&&\alpha\equiv F_X^0-X_0 F_{XX}^0\;,\;\;\beta\equiv3F_{XX}^0+2F_Y^0+2F_Z^0\;\nonumber\\
&&\gamma\equiv F_Y^0+4F_Z^0\;.\label{coefficients}\eea 

Now, since we will be interested in theories of the kind $F(X,Y,Z)=F(X,{\cal G})=X+f({\cal G})$, then it will be convenient to make the following replacements in the above expansion:

\bea &&F_X^0\rightarrow 1+f_{\cal G}^0{\cal G}_X^0,\;F_Y^0\rightarrow f_{\cal G}^0{\cal G}_Y^0,\;F_Z^0\rightarrow f_{\cal G}^0{\cal G}_Z^0,\nonumber\\&&F_{XX}^0\rightarrow f_{{\cal G}{\cal G}}^0({\cal G}_X^0)^2+f_{\cal G}^0{\cal G}_{XX}^0,\label{replacements}\eea where, as before, the suffix ''$0$'' means that the given magnitude is evaluated at background curvature $X_0$, and we have taken into account the fact that, for $F(X,{\cal G})$-theories of the form: $F(X,{\cal G})=X+f({\cal G})$, $F_X=1$, $F_{XX}=0$, and $F_{\cal G}=f_{\cal G}$. Besides, since ${\cal G}_X=2X$, ${\cal G}_{XX}=2$, ${\cal G}_Y=-4$, and ${\cal G}_Z=1$, then we can write the expansion of the above $F(X,{\cal G})$-theories around vacuum of constant curvature $X_0$, in the following form:

\bea &&F(X,{\cal G})=f_0-\frac{1}{6}X_0^2 f_{\cal G}^0+2X_0^4 f_{{\cal G}{\cal G}}^0\nonumber\\
&&\;\;\;\;\;\;\;\;\;\;+(1-4X_0^3 f_{{\cal G}{\cal G}}^0)X+2X_0^2 f_{{\cal G}{\cal G}}^0 X^2+{\cal O}(3).\nonumber\eea The linearized function $F(X,{\cal G})$ can be further written in a form that is convenient to compare it with known results when the metric is perturbed around the vacuum solution \cite{stelle,ovrut,solganik,chiba}:

\be F(X,{\cal G})=\alpha\left(-2\Lambda+X+\frac{1}{6m^2}X^2\right),\label{linearized}\ee where we have defined the effective cosmological constant 

\be -2\Lambda\equiv\frac{f_0-X_0^2 f_{\cal G}^0/6+2X_0^4 f_{{\cal G}{\cal G}}^0}{1-4X_0^3 f_{{\cal G}{\cal G}}^0},\label{L}\ee and 

\be m^2\equiv\frac{1-4X_0^3 f_{{\cal G}{\cal G}}^0}{12 X_0^2 f_{{\cal G}{\cal G}}^0},\label{mass}\ee is the mass squared of the spin-0 perturbation propagating in the background space. 

As seen from (\ref{linearized}), the linearized $F(X,{\cal G})$-theory can be recast into the form of an $f(X)$-theory (see also \cite{quiros-urena}), so that, at linearized level, it is also Ostrogradski-stable. In what follows, under the term ''Ostrogradski-stable'', just linear Ostrogradski-stability is to be meant. 

The constant \be\alpha\equiv 1-4X_0^3 f_{{\cal G}{\cal G}}^0,\label{alpha}\ee in (\ref{linearized}) plays an important role since it multiplies the gravitational action, thus modifying the gravitational coupling. In fact, by substituting the linearized function (\ref{linearized})  back into the action (\ref{action}) one gets that $$S_g=\frac{\alpha}{2\kappa^2}\int d^4x\sqrt{|g|}\left(-2\Lambda+X+\frac{1}{6m^2}X^2\right)\;,$$ so that the effective gravitational coupling is given by 

\be 8\pi G_{eff}=\frac{\kappa^2}{\alpha}=\frac{\kappa^2}{1-4X_0^3 f_{{\cal G}{\cal G}}^0}.\label{eff grav coupling}\ee If a given $F(X,{\cal G})$-theory fails to generate a non-negative effective gravitational coupling, then the graviton is a negative norm state (a ghost), and the theory has to be rule out as unphysical. Within $f(X)$-theories fulfillment of this requirement (non-negativity of $G_{eff}$) is known as ''linear stability''.

We have to differentiate $G_{eff}$ from the effective Newton's ''constant'' $8\pi G_N^{eff}=\kappa^2/F_X$, which is multiplying the right-hand-side (RHS) of the Einstein's field equations:

\be G_{\mu\nu}=8\pi G_N^{eff}(T_{\mu\nu}^{(m)}+T_{\mu\nu}^{cur})\;,\label{feqs}\ee which can be derived from (\ref{action}) by varying with respect to the metric \cite{carroll}. In (\ref{feqs}), $T_{\mu\nu}^{(m)}$ is the stress-energy tensor for matter, while

\bea \kappa^2 T_{\mu\nu}^{cur}=\frac{1}{2}g_{\mu\nu}(F-X F_X)-(g_{\mu\nu}\Box-\nabla_\mu\nabla_\nu) F_X\nonumber\\
-2(F_Y R_{\;\;\mu}^{\sigma}R_{\sigma\nu}+F_Z R_{\lambda\sigma\rho\mu}R^{\lambda\sigma\rho}_{\;\;\;\;\;\;\nu})-g_{\mu\nu}\nabla_\sigma\nabla_\lambda(F_Y R^{\sigma\lambda})\nonumber\\-\Box(F_Y R_{\mu\nu})+2\nabla_\sigma\nabla_\lambda(F_Y R^\sigma_{\;\;(\mu}R^\lambda_{\;\;\nu)}+2F_Z R^{\sigma\;\;\;\;\;\lambda}_{\;(\mu\nu)}),\nonumber\eea where $\Box\equiv g^{\mu\nu}\nabla_\mu\nabla_\nu$ is the D'Lambertian, and we have used the usual representation for symmetrization: $T_{(\mu\nu)}=(T_{\mu\nu}+T_{\nu\mu})/2$. This form of writing of the field equations is particularly useful for cosmological applications. The trace of equation (\ref{feqs}) generates an additional constraint on the curvature, which, for vacuum spaces ($T^{(m)}=0$) of constant curvature can be written as:

\be 2F_0-X_0 F_X^0-2Y_0 F_Y^0-2Z_0 F_Z^0=0.\label{trace}\ee As a check of consistency, notice that by taking into account the trace equation (\ref{trace}), which when evaluated at background constant curvature $X_0$, in terms of the ${\cal G}$-derivatives of $f({\cal G})$, amounts to:

\be X_0+2f_0-\frac{1}{3}X_0^2 f_{\cal G}^0=0\;,\label{trace'}\ee then the effective cosmological constant computed with the help of (\ref{L}), yields $\Lambda=X_0/4$, as it should be for a background space of constant curvature.

If in equations (\ref{feqs}) one replaces $F_X\rightarrow 1+2X f_{\cal G}$, etc., then, one can write the effective Newton's constant in the following form:

\be 8\pi G_N^{eff}=\frac{\kappa^2}{1+2X f_{\cal G}}.\label{eff newton constant}\ee Note that, since equations (\ref{feqs}) hold in general, the magnitudes in (\ref{eff newton constant}) have not been evaluated at constant curvature. However, in order to compare with other bounds, we can indeed evaluate at constant $X=X_0$. 

Another kind of catastrophic instability, the so called ``Ricci instability'' \cite{faraonirev}, also known as Dolgov-Kawasaki instability \cite{kawasaki} (see also \cite{odintsov1,faraoni2007}), is associated, precisely, with the behavior of the effective Newton's constant $G_N^{eff}$. Although the issue has been well explored within $f(X)$-theories, for $F(X,Y,Z)$-theories a detailed study of this issue is lacking. While in $f(X)$-theories Ricci stability is closely related to the absence of spin-0 tachyon instability, for $F(X,Y,Z)$ gravity theories these are separate requirements. To understand the issue one might rely on the following reasoning line. Following the arguments in \cite{faraonirev}, suppose that the magnitude of the effective Newton's constant, which variation is governed by $$\frac{dG_N^{eff}}{dX}=-\frac{\kappa^2}{8\pi F_X^2}F_{XX}=-\frac{2\kappa^2(f_{\cal G}+2X^2 f_{\cal GG})}{8\pi(1+2X f_{\cal G})^2}>0\;,$$ increases as the curvature $X$ increases. Hence, at large curvature the effect of gravity becomes stronger. Additionally, increasing $X$ generates larger values of the curvature through the trace equation (\ref{trace}). The combination of the above mechanisms conspires to destabilize the theory: a (initially) small curvature grows without limits and the system runs away. In order to avoid this kind of unwanted instability one has to require that $$\frac{dG_N^{eff}}{dX}\leq 0\;\Rightarrow\;f_{\cal G}+2X^2 f_{\cal GG}\geq 0\;.$$ Theories that meet the above bound are said to be Ricci stable \cite{faraonirev,faraoni2007}. Since the mass squared of the scalar degree of freedom in linearized $f(X)$-theories is given by $$m_\vphi^2=\frac{f_X^0-X_0 f_{XX}^0}{3f_{XX}^0}\;,$$ and since linear stability of the scalar perturbation requires that $f_X^0-X_0 f_{XX}^0\geq 0$ \cite{starobinski}, hence, absence of spin-0 tachyon instability within $f(X)$-theories requires that $f_{XX}^0>0$ (the theory has to be Ricci stable). This is why, within $f(X)$-theories Ricci stability is closely related with absence of a scalar ghost.\footnote{Used in this context the meaning of ''ghost'' state is misleading, since, in field theory a ghost is usually associated with a state of negative ''kinetic'' energy (otherwise a state of negative norm), while a state with negative mass squared is usually associated with a tachyon degree of freedom.} When dealing with $F(X,Y,Z)$-theories the situation is a bit different. Actually, due to the definition of the mass squared of the spin-0 perturbation in Eq. (\ref{mass}), the theory can be Ricci stable $f_{\cal G}^0+2X_0^2 f_{\cal GG}^0\geq 0$, and, at the same time, the scalar perturbation can be linearly stable ($G_{eff}\geq 0$) $1-4X_0^3 f_{\cal GG}^0\geq 0$, and yet it can develop a tachyon instability as long as $f_{\cal GG}^0<0$. 

The instability caused by the presence of a spin-0 tachyon degree of freedom, can be catastrophic either since the time scale for this instability to manifest is estimated to be of the order of $\propto m_0^{-1}$, and it might develop as quickly as in a time $\sim 10^{-26}\;s$ \cite{kawasaki}.

Summarizing our discussion. The relevant stability requirements imposed on theoretically consistent quadratic modifications of gravity of the kind $F(X,{\cal G})=X+f({\cal G})$, are the following \cite{quiros-urena}:

\begin{itemize}

\item{Ostrogradski Stability:} The linearized $F(X,{\cal G})$-theory should be expressible as an equivalent $f(X)$-theory.\footnote{Stated in this form Ostrogradski stability implies also, absence of spin-2 Weyl ghost propagating modes.} Theories of the kind $F(X,{\cal G})$ are obviously spin-2 ghost-free, and their linearizations can be written in the form of an $f(X)$-theory. Hence, the models of interest here (\ref{powers}-\ref{zcs C}), are free of the Ostrogradski instability, at least at the linearized level. 

\item{Non-negativity of the (inverse) Effective Gravitational Coupling:} $$\frac{\alpha}{\kappa^2}=\frac{1-4X_0^3 f_{\cal GG}^0}{\kappa^2}\geq 0\;.$$ Fulfillment of this constraint warrants that the graviton is not a ghost. 

\item{Absence of Tachyon Instability:} $$m^2=\frac{1-4X_0^3 f_{\cal GG}^0}{12 X_0^2 f_{\cal GG}^0}\geq 0\;.$$ Once non-negativity of the effective gravitational coupling is verified, the requirement of absence of spin-0 tachyon instability amounts just to $f_{\cal GG}^0>0$. Fulfillment of the latter constraint means that the scalar degree of freedom is not a destabilizing tachyon.

\item{Ricci Stability:} $$f_{\cal G}^0+2X_0^2 f_{\cal GG}^0\geq 0\;.$$ This requirement guarantees that the graviton does not become a ghost.

\end{itemize} 

Note that the bound established on viable quadratic theories of gravity by the dynamical systems study, according to which stability of the de Sitter point requires that $f_{\cal GG}\geq 0$ \cite{dft}, is consistent with the requirement of absence of tachyon instability (third item above) once non-negativity of the effective gravitational coupling is achieved. As already seen the latter is an independent requirement. 

We want to recall that spin-2 ghost-free theories are also Ostrogradski stable in the sense discussed before. Notice that for an $f(X)$-theory, the last three requirements are not independent: Non-negativity of $G_{eff}$ (linear stability) and Ricci stability, together imply that the scalar mode is not a tachyon. 

Now we are in position to discuss the stability of theories given by (\ref{powers}-\ref{zcs C}), regarding the last three requirements above.

\section{Instabilities in the Model $F(X,{\cal G})=X+\mu{\cal G}^n$}\label{power-law models}

The function $F(X,{\cal G})$ in the action (\ref{powers}) is $$F(X,{\cal G})=X+\mu{\cal G}^n\;,$$ where, for generality of the analysis we keep arbitrary the sign of the constants $\mu$ and $n$, i.  e., $\mu\in\Re$, $n\in\Re$. The derived magnitudes $f_{\cal G}=n\mu{\cal G}^{n-1}$, $f_{\cal GG}=n(n-1)\mu{\cal G}^{n-2}$, when evaluated at the background value $X_0$ (recall that ${\cal G}_0=X_0^2/6$) are given by:

\be f_{\cal G}^0=\frac{n\mu}{6^{n-1}}X_0^{2n-2},\;f_{\cal GG}^0=\frac{6n(n-1)\mu}{6^{n-1}} X_0^{2n-4}.\label{vacuum}\ee 

The value of the background's constant curvature $X_0$ is determined by means of the trace equation (\ref{trace'}), which, in the present case can be written as $$X_0+2f_0-\frac{X_0^2}{3} f_{\cal G}^0=\frac{3X_0}{6^n}\left[6^n+2\mu(1-n)X_0^{2n-1}\right]=0\;.$$ As seen, the latter requirement amounts to an algebraic equation\footnote{This is true only for power-law functions of the GB invariant.} which can be solved, in principle, to get $X_0$ as a function of the overall parameters: $$X_0^{2n-1}=\frac{6^n}{2(n-1)\mu}\;.$$ Taking into account the above relationships we get that:

\bea &&1-4X_0^3 f_{\cal GG}^0=1-72n,\;\;f_{\cal GG}^0=\frac{18n}{X_0^3},\nonumber\\
&&f_{\cal G}^0+2X_0^2f_{\cal GG}^0=\frac{3n(12n-11)}{(n-1)X_0}.\nonumber\eea 

Note that the stability bounds discussed above for the model (\ref{powers}), do not depend on the parameter $\mu$. Therefore, the sign of the quadratic modification of GR: $f({\cal G})\propto{\cal G}^n$, does not play role in the stability of its linearization around maximally symmetric background spaces of constant curvature. Let us write the relevant stability criteria for the model in terms of requirements on the parameter $n$:

\begin{enumerate}

\item{Non-negativity of the Gravitational Coupling:} $$1-72n\geq 0,\;\Rightarrow\;n\leq\frac{1}{72}\;.$$

\item{Absence of Tachyon Instability:} $$1-72n\geq0,\;n>0\;.$$ 

\item{Ricci Stability:} $$n(12n-11)\geq 0,\;\Rightarrow\;n\geq\frac{11}{12}\;.$$

\end{enumerate}

It is unfortunate for the quadratic theories that are power-law in the GB invariant, that the above requirements can not be jointly met, so that these theories are not theoretically consistent in the sense discussed above. For negative $n<0$, in particular, the bounds 1-3 reveal that, while the effective gravitational coupling $G_{eff}$ can be a positive quantity, as long as, for negative $n=-\bar n$ ($\bar n>0$), $1+72\bar n\geq 0$ always, the theory (\ref{powers}) develops a tachyon instability since the mass squared of the propagating scalar degree of freedom is a negative quantity (recall that we are considering just positive curvature backgrounds): $$m^2=-\frac{1+72\bar n}{216\bar n}X_0<0\;.$$ As already discussed, this is a serious objection against the quadratic modification of gravity depicted by the action (\ref{powers}) with positive and negative $n$-s. For $f({\cal G})$ in (\ref{powers}) with inverse powers of the GB invariant (negative $n<0$), the model considered here is a particular (spin-2 ghost-free) case of the model studied in Ref. \cite{navarro,akoleyen}. 

We want to comment that, for positive $n>0$, the constraints established on (\ref{powers}) by consistency with solar-system test and with present cosmic speedup lead to $n\lesssim 0.074$ \cite{davis}. This is an example of a toy model that passes the cosmological tests while being theoretically inconsistent. 

Our previous results also apply to the more general model $f({\cal G})=\mu |{\cal G}|^n$ of Ref. \cite{odintsov2}, since, in this case, $$f_{\cal G}=n\mu {\cal G}|{\cal G}|^{n-2},\;f_{\cal GG}=n(n-1)\mu|{\cal G}|^{n-2}\;,$$ so that, at constant curvature, ${\cal G}_0=|{\cal G}|=X_0^2/6$, is always a positive magnitude.

\section{Instabilities in the Toy Models of Ref. \cite{dft}}\label{dft models}

The toy models (\ref{dft A}-\ref{dft C}) were proposed in Ref. \cite{dft} on the basis of several mathematical requirements on the function $f({\cal G})$ that make them cosmologically viable models. The most crucial condition to be satisfied is $f_{\cal GG} > 0$, which is required to ensure the stability of a late-time de-Sitter solution, as well as the existence of standard radiation/matter dominated epochs. The explicit $f({\cal G})$ models (\ref{dft A}-\ref{dft C}) show a stage of cosmic acceleration followed by the matter era. Although these models have been shown to be free of ghost instabilities, on the light of the discussion in the former sections we shall explore them with regard to the stability of vacuum solutions of constant curvature, by focusing on the requirements of Ricci stability and of non-negativity of the effective gravitational coupling.

\subsection{Toy model given by equation (\ref{dft A})}

The first check will be to ask for positivity of the effective gravitational coupling $8\pi G_{eff}=\kappa^2\alpha^{-1}$, i. e., $$1-4X_0^3 f_{\cal GG}^0\geq 0\;.$$ Let us write the function $f({\cal G})$ given by equation (\ref{dft A}), in the following form: $$f(\xi)=\lambda\sqrt\beta[\xi\arctan\xi-\frac{1}{2}\ln(1+\xi^2)-\alpha]\;,$$ where we have introduced a new field variable $\xi={\cal G}/\beta$  ($\xi_0=X_0^2/6\beta$), so that $f_\xi=\beta f_{\cal G}$, and $f_{\xi\xi}=\beta^2 f_{\cal GG}$. Hence, $$f_\xi=\lambda\sqrt\beta\arctan\xi,\;f_{\xi\xi}=\frac{\lambda\sqrt\beta}{1+\xi^2}\;.$$ The requirement of non-negativity of $G_{eff}$ becomes into the following inequality: $$X_0^4-144\lambda\sqrt\beta X_0^3+36\beta^2\geq 0\;.$$ Since $\lambda$ (also $\beta$, since the square root $\sqrt\beta$ has to be real) is non-negative, hence there is an interval in the values of the curvature $X_0$ lying between the smaller and larger positive roots of the polynomial equation $X_0^4-144\lambda\sqrt\beta X_0^3+36\beta^2=0$, where the above bound is not satisfied. Here a cautionary note is necessary: the value of the background constant curvature $X_0$ can no be chosen at will, it is completely determined in terms of the free constant parameters $\lambda$, $\beta$, $\alpha$, by the trace equation (\ref{trace'}) which, in the present case, amounts to the following non-algebraic equation: $$\ln\left(1+\frac{X_0^4}{36\beta^2}\right)=\frac{X_0}{\lambda\sqrt\beta}-2\alpha\;.$$ Since the LHS of this equation is always positive, then, any root $X_0^*$ of this equation will obey $X_0^*> 2\alpha\lambda\sqrt\beta$.

That the model does not develop gravitational ghosts is evident from the following inequality, expressing the requirement of Ricci stability ($f_{\cal G}^0+2X_0^2 f_{\cal GG}^0\geq 0$): $$\arctan\left(\frac{X_0^2}{6\beta}\right)+\frac{72\beta X_0^2}{36\beta^2+X_0^4}\geq 0\;,$$ which is always met. The model is also spin-0 tachyon-free, since the constraint $f_{\cal GG}>0$ was one of the requirements imposed in \cite{dft} to obtain the viable models (\ref{dft A}-\ref{dft C}).

\subsection{Toy model given by equation (\ref{dft B})}

In this case, since $$f({\cal G})=\frac{\lambda}{\sqrt\beta}{\cal G}\arctan\left(\frac{\cal G}{\beta}\right)-\alpha\lambda\sqrt\beta\;,$$ or, in terms of the variable $\xi={\cal G}/\beta$ $$f(\xi)=\lambda\sqrt\beta(\xi\arctan\xi-\alpha)\;,$$ then $$f_\xi=\lambda\sqrt\beta\left(\arctan\xi+\frac{\xi}{1+\xi^2}\right),\;f_{\xi\xi}=\frac{2\lambda\sqrt\beta}{(1+\xi^2)^2}\;.$$ The trace equation (\ref{trace'}) leads to the following 5th-order algebraic equation to determine the background curvature $X_0$ as function of the free parameters $\alpha$, $\lambda$, $\beta$, $$\frac{X_0^4}{36\beta^2+X_0^4}=\frac{X_0}{2\lambda\sqrt\beta}-\alpha\;.$$ Non-negativity of the effective gravitational coupling can be expressed in the form of the following inequality: $$X_0^5-36\cdot 143\beta^2X_0+8\cdot 36^2\alpha\lambda\beta^{5/2}\geq 0\;,$$ where the trace equation has been taken into account. Extracting useful qualitative information from the above inequality is very difficult. Absence of Ricci instability ($f_{\cal G}^0+2X_0^2 f_{\cal GG}^0\geq 0$) is guaranteed since ($\beta>0$): $$\arctan\left(\frac{X_0^2}{6\beta}\right)+\frac{X_0^2(25+X_0^4/36\beta^2)}{6\beta(1+X_0^4/36\beta^2)^2}\geq 0\;.$$

\subsection{Toy model given by equation (\ref{dft C})}

According to (\ref{dft C}) we have: $$f({\cal G})=\lambda\sqrt\beta\ln\left[\cosh\left(\frac{\cal G}{\beta}\right)\right]-\alpha\lambda\sqrt\beta\;.$$ For simplicity of writing let us to introduce the following parameters: $$\xi\equiv\frac{{\cal G}_0}{\beta}=\frac{X_0^2}{6\beta}\;,\;\;W_0\equiv\frac{\sinh\xi}{\cosh\xi}\;,$$ then, non-negativity of the effective gravitational coupling $G_{eff}$ and Ricci stability requirement, can be written in the form of the following inequalities: $$X_0^3(1-W_0^2)\leq\frac{\beta^{3/2}}{4\lambda},\;\beta W_0+2X_0^2(1-W_0^2)\geq 0\;,$$ respectively. While the latter requirement is always fulfilled, non-negativity of $G_{eff}$ requires a careful examination in parameter space due to the bound $X_0^3(1-W_0^2)\leq\beta^{3/2}/4\lambda$ (the trace equation has to be considered).

Although the models (\ref{dft A}-\ref{dft B}) of reference \cite{dft} seem to be theoretically consistent also from the point of view of stability of vacuum, the cosmological constant term may be needed in general for the cosmological viability of these models ($\alpha\neq 0$). As the authors of \cite{dft} recognize, the need of such a constant term makes the models (\ref{dft A}-\ref{dft B}) less attractive from a theoretical point of view.

\section{Instabilities in the Toy Models of Ref. \cite{zcs}}\label{zcs models}

Models (\ref{zcs A}-\ref{zcs C}) where explored from the point of view of their asymptotic properties in Ref. \cite{zcs}. The authors demonstrated that cosmologically viable trajectories in the phase space of the model can be found. In particular a stable de Sitter point, corresponding to late-time cosmological evolution -- co-existing together with phantom like stage --, is found. The behavior of the relevant cosmological parameters of observational interest is quite similar to the one obtained in the $\Lambda$CDM-model. All of these models produce an elongated matter dominated epoch followed by a radiation dominated epoch \cite{zcs}. On the basis of their study the authors concluded that the toy models (\ref{zcs A}-\ref{zcs B}) are cosmologically viable. Here we aim at exploring the above toy models from the perspective of the vacuum structure of their linearizations in the sense discussed in the former sections.

\subsection{Model given by equation (\ref{zcs A})}

This toy model is given by the function: $$f({\cal G})=\alpha {\cal G}^{1/2}+\beta{\cal G}^{1/4}\;,$$ where $\alpha$ and $\beta$ are arbitrary constants. In this case the trace equation (\ref{trace'}) is easily solved to yield $$\sqrt{X_0}=-\frac{6^{3/4}\beta}{4(1+\alpha/\sqrt 6)}\;.$$ Hence, for instance $${\cal G}_0^{-1/2}=\frac{4(\sqrt{6}+\alpha)^2}{9\beta^2},\;{\cal G}_0^{-1/4}=-\frac{2(\sqrt{6}+\alpha)}{3\beta}\;,$$ etc. Besides $$f_{\cal G}^0=\frac{2^2}{3^3\beta^2}(\alpha+\sqrt 6)^2(\alpha-\sqrt{3/2})\;,$$ and $$f_{\cal GG}^0=-\frac{2^3}{3^5\beta^6}(\alpha+\sqrt 6)^6(\alpha-\sqrt{2/3})\;.$$ Non-negativity of the effective gravitational constant $G_{eff}$: $1-4X_0^3f_{\cal GG}^0\geq 0$, is achieved whenever $$\alpha\geq\frac{17}{18}\sqrt\frac{2}{3}\;,$$ while, absence of tachyon instability ($G_{eff}\geq 0$, $f_{\cal GG}^0>0$), imposes the following constraint: $$-(\alpha+\sqrt 6)^2(\alpha-\sqrt{2/3})>0\;\Rightarrow\;\frac{17}{18}\sqrt\frac{2}{3}\leq\alpha<\sqrt\frac{2}{3}\;.$$ Ricci stability: $f_{\cal G}^0+2X_0^2 f_{\cal GG}^0\geq 0$, requires that $$\frac{2}{3^3\beta^2}(\alpha+\sqrt 6)^2(8\sqrt{6}-25\alpha)\geq 0\;\Rightarrow\;\alpha\leq\frac{24}{25}\sqrt\frac{2}{3}\;.$$ We see that Ricci stability, non-negativity of the effective gravitational coupling, and absence of tachyon instability, can be simultaneously fulfilled only if $$\frac{17}{18}\sqrt\frac{2}{3}\leq\alpha\leq\frac{24}{25}\sqrt\frac{2}{3}\;,$$ or if: $0.7711\leq\alpha\leq 0.7838$. By comparing this result with the best-fit value of $\alpha$ reported in Tab. I of reference \cite{cosmo tests}: $$\alpha=0.00084_{-0.01632}^{+0.00016}\;,$$ one sees that this value (including error bars) does not fit into the above narrow interval where the model is theoretically consistent.

The conclusions drawn from the above results suggest that cosmological viability of the model (\ref{zcs A}) is in serious trouble if one requires also its theoretical consistency.

\subsection{Model given by equation (\ref{zcs B})}

Another of the toy models proposed in \cite{zcs}, whose cosmological viability was demonstrated in Ref. \cite{cosmo tests} due to its good fitting to combined cosmological data sets, is given by the following function: $$f({\cal G})=\alpha({\cal G}^{3/4}-\beta)^{2/3}\;.$$ Hence, $$f_{\cal G}=\frac{\alpha}{2}({\cal G}^{3/4}-\beta)^{-1/3}{\cal G}^{-1/4},\;$$ $$f_{\cal GG}=-\frac{\alpha}{4}({\cal G}^{3/4}-\beta)^{-4/3}({\cal G}^{3/4}-\beta/2){\cal G}^{-5/4}\;.$$ The trace equation for this model (considering, as we have done, a vacuum background of constant curvature where ${\cal G}_0=X_0^2/6$) can be written as: $$(1-\xi)+\frac{\sqrt 6}{2\alpha}(1-\xi)^{1/3}-\frac{1}{2}=0\;,$$ where we have defined $\xi\equiv\beta/{\cal G}_0^{3/4}=6^{3/4}\beta/X_0^{3/2}$. The only real root of the latter equation is given by $$\xi=1-r_*^3\;,\;\;r_*=\left(\frac{l}{36}\right)^{1/3}-\frac{\sqrt{6}/\alpha}{(6l)^{1/3}}\;,$$ with $l\equiv 9+\sqrt{81+36\sqrt{6}/\alpha^3}$ (since $l$ is real, then, $\alpha\leq-1.02872$). In terms of the variable $\xi$, taking into account the trace equation above, one has that: $$f_{\cal G}^0=\frac{\sqrt 6}{2}\left(\frac{\xi}{\beta}\right)^{2/3}\frac{1}{2\xi-1}\;,$$ and $$f_{\cal GG}^0=\frac{\sqrt 6}{8\beta^2}\frac{\xi^2(2-\xi)}{(2\xi-1)(\xi-1)}\;.$$ Hence, non-negativity of $G_{eff}$ will imply that $$1-4X_0^3 f_{\cal GG}^0=\frac{2\xi^2+15\xi-35}{(2\xi-1)(\xi-1)}\geq 0\;,$$ which leads to $$-\infty<\xi\leq -9.3681\;;\;\;0.5<\xi<1\;,$$ and, also $$1.8681\leq\xi<\infty\;.$$ Absence of tachyon instability can be written as $$f_{\cal GG}^0=\frac{\sqrt 6}{8\beta^2}\frac{\xi^2(2-\xi)}{(2\xi-1)(\xi-1)}\geq 0\;\Rightarrow\;\xi<1/2,\;1<\xi\leq 2\;.$$ Additionally, Ricci stability bound is fulfilled if $$\left(\frac{\xi}{\beta}\right)^{2/3}\frac{5-2\xi}{(2\xi-1)(\xi-1)}\geq 0\;\Rightarrow\;\xi<1/2,\;1<\xi\leq 5/2\;.$$ Combining the above stability requirements for theoretical consistency of the model (\ref{zcs B}), yields to the following constraint on the parameter $\xi$, that has to be fulfilled by the model if it is to be considered as theoretically consistent: $$\beta>0\;;\;\;1.8681\leq\xi\leq 2\;,$$ while, for negative $\beta$-s: $$\beta<0\;;\;\;-\infty<\xi\leq -9.3681\;.$$ If one substitutes these limits back into the trace equation, one gets the corresponding allowed ranges of the parameter $\alpha$: $$\beta>0\;;\;\;-1.6984\leq\alpha\leq -1.6706\;,$$ and $$\beta<0\;;\;\;-1.9764\leq\alpha\leq-1.02872\;.$$ On the other hand, in Tab. I of Ref. \cite{cosmo tests} the reported best-fit value of the parameter $\alpha$: $$\alpha=-0.00014^{+0.000083}_{-0.001439}\;,$$ so that, considering error bars, the model is only marginally consistent.

\subsection{Model given by equation (\ref{zcs C})}

The last toy model we will explore here is the one given by the following function: $$f({\cal G})=\alpha\sqrt{\cal G}\exp\left(\frac{\beta}{\cal G}\right)\;.$$ It will be useful to introduce the parameter $\xi\equiv\beta/{\cal G}_0$, which, depending on the sign of $\beta$ can be either positive or negative. We have $$f_0=\alpha\sqrt\frac{\beta}{\xi}\;e^\xi,\;f_{\cal G}^0=\frac{\alpha}{2}\sqrt\frac{\xi}{\beta}\;e^\xi(1-2\xi)\;,$$ and $$f_{\cal GG}^0=-\frac{\alpha}{4}\frac{\xi^{3/2}}{\beta^{3/2}}\;e^\xi(1-4\xi-4\xi^2)\;.$$ In terms of these parameters the trace equation (\ref{trace'}) can be written as $$e^{-\xi}=-\frac{\alpha}{\sqrt 6}(1+2\xi)\;.$$ This equation will be used to remove the exponential from the expressions below. 

Non-negativity of $G_{eff}$ can be written as the following constraint $$\frac{4\xi^2+\frac{73}{18}\xi-\frac{35}{36}}{1+2\xi}\geq 0\;,$$ which can be solved to yield: $$-1.2141\leq\xi<-0.5\;;\;\;0.2002\leq\xi<\infty\;.$$ Once $G_{eff}$ is checked to be non-negative, absence of spin-0 tachyon mode amounts to $f_{\cal GG}^0>0$: $$\frac{1-4\xi-4\xi^2}{1+2\xi}>0\;.$$ Hence: $$-\infty<\xi<-1.2071\;;\;\;-0.5<\xi<0.2071\;.$$ Joint fulfillment of the above requirements, meaning non-negative $G_{eff}$ and absence of scalar tachyon, is possible only if the parameter $\xi$ falls within the following, very narrow intervals: $$-1.2141\leq\xi<-1.2071\;;\;\;0.2002\leq\xi<0.2071\;.$$ 

Ricci stability $$f_{\cal G}^0+2X_0^2 f_{\cal GG}^0\geq 0\;\Rightarrow\;\frac{5-22\xi-24\xi^2}{1+2\xi}\geq 0\;,$$ imposes additional constraints: $$-\infty<\xi\leq-1.1052\;;\;\;-0.5<\xi\leq 0.1885\;.$$ The latter ranges match with the ones where joint fulfillment of $G_{eff}\geq 0$ and of $f_{\cal GG}^0>0$, is allowed, only if $-1.2141\leq\xi<-1.2071$, or, in terms of the original parameters: $$-4.9706\beta<X_0^2\leq-4.9419\beta\;,$$ which forces the parameter $\beta$ to be negative. Besides, if one substitutes the values of $\xi$ within the above range, back into the trace equation, one gets that the parameter $\alpha$ takes values within the following interval $$5.7752\leq\alpha<5.7916\;,$$ which, including error bars, is inconsistent with the best-fit $\alpha$-value $$\alpha=-0.00012_{-0.00295}^{+0.000004}\;,$$ reported in Tab. I of reference \cite{cosmo tests}. Therefore, observational viability of this model is obviously in conflict with its theoretical consistency.

\section{Discussion}\label{discussion}

A careful study of the stability of higher-order modifications of gravity has been proved to be useful in ruling out theoretically inconsistent models. This is particularly true for $f(X)$-theories. It has been demonstrated, in particular, that these theories are equivalent to scalar-tensor gravity, with a curvature-dependent mass squared of the scalar field (see the review \cite{faraonirev} and references therein). The dependence on the curvature makes it possible to explain the modifications in the large scale caused by the scalar mode: very light mass of the scalar implies that the modification of gravity has implications for cosmology, while, heavy mass in the neighborhood of galaxies and star systems, makes the scalar mode to be shielded at small scales thus rendering Newton's law of gravity valid. When dealing with arbitrary quadratic modifications of gravity of the form $F(X,Y,Z)$, the situation is much more complicated. In this case the original theory is shown to be equivalent to a multi(quadruple)-scalar-tensor theory with the scalars coupled to the different curvature invariants \cite{chiba}. This makes the equivalence useless to study the field content of the theory as well as to study observational constraints on it. In order to study such fundamental issues one, usually, limits oneself to considering excitations around a vacuum state (a maximally symmetric spacetime with a constant curvature) \cite{ovrut}. First one expands the action around spaces of constant curvature, and then one incorporates linear perturbations of the metric around vacuum solutions. The spectrum of perturbations is then revealed. For general theories $F(X,Y,Z)$ there arise 8th degrees of freedom: two associated with the spin-2 (massless) graviton, one with the spin-0 scalar mode, and five with the spin-2 massive Weyl state \cite{stelle,ovrut,solganik,chiba}. The latter happens to be a ghost due to the negative sign of the Weyl term after the linearization. Much information about the original (unperturbed) theory can be extracted from the properties of the above spectrum of perturbations around vacuum background spaces. Actually, if one wants to have a consistent expansion of the theory around such vacuum states, then one has to require, in the linearized theory: i) the effective gravitational coupling to be non-negative, so that the massless graviton is a normalizable state, ii) the mass squared of the scalar propagating mode to be non-negative since, otherwise, the spin-0 mode would be a tachyon state leading to disastrous destabilization of the vacuum, and iii) the negative mass squared of the spin-2 Weyl mode to be large enough as to decouple from the linearized spectrum (no unphysical Weyl ghost, or ''poltergeist''). The last requirement (spin-2 ghost free theory) is achieved if one considers particular quadratic modifications of the kind studied here: $F(X,{\cal G})=X+f({\cal G})$. At the linearized level these are scalar-tensor theories of gravity.

Following discussions within $f(X)$-theories, one might invoke other instabilities that are of importance in cosmological applications. In particular, the so called Ricci, or also, Dolgov-Kawasaki instability, associated with the fact that the graviton might become a ghost if the curvature increases \cite{odintsov1,faraonirev,faraoni2007,kawasaki}. As it is usually done in cosmological applications of $f(X)$-theories, working with $F(X,Y,Z)$ models one can write the LHS of the field equations in the form of standard Einstein's equation, with a non-standard RHS. In this case, the effective gravitational (Newton's) constant $G_N^{eff}$ depends on the curvature -- strictly speaking, it depends on the $X$-derivative of the function $F(X,Y,Z)$: $G_N^{eff}\propto F_X^{-1}$ --, so that it varies with the course of the cosmic expansion. It might happen that, as the cosmic expansion proceeds the value of $G_N^{eff}$ increases. This would has a disastrous impact on the theory, since the curvature might grow without limits \cite{faraonirev,faraoni2007}. Therefore, one has to require the $X$-derivative of the effective Newton's constant to be non-positive: $dG_N^{eff}/dX\leq 0$, which means that $G_N^{eff}$ is a non-increasing function of the background curvature. This will warrant that the graviton does not become a ghost as the background curvature grows.

A combination of the above requirements (non-negativity of the effective gravitational coupling, Ricci stability, absence of scalar tachyon) might be very useful to rule out apparently viable quadratic theories of gravity. In this paper we have checked several such cosmologically viable theories \cite{navarro,odintsov2,dft,zcs,cosmo tests}, where the function $F(X,Y,Z)=F(X,{\cal G})=X+f(\cal G)$, regarding the above discussed stability tests. It arose that some of these ''viable'' models did not pass the stability tests considered here. In particular, if one takes into account the best-fit values that make possible to talk about cosmological viability of some of the models, one sees that, either the models clearly fail to meet the stability requirements, or they are on the edge of (not) being theoretically consistent. 

It has been demonstrated that theories based on Lagrangians which contain powers of the GB invariant (for negative powers $n<0$, these are a particular case of the model proposed in Ref. \cite{navarro}), are Ricci unstable and, also, develop scalar tachyons. Ricci and tachyon instabilities combined, provoke gross destabilizing effects with catastrophic consequences for the theory. This result applies also to models of the kind $f({\cal G})=\mu|{\cal G}|^n$, thoroughly explored in Ref. \cite{odintsov2}. A careful study of the models (\ref{zcs A}-\ref{zcs C}) \cite{zcs}, reveals that, when taking into account the reported best-fit values of the parameters of the models, the stability criteria are either clearly not met (models (\ref{zcs A}) and (\ref{zcs C})), or are at the edge of being not met (model (\ref{zcs B})). While the situation with models (\ref{dft A}-\ref{dft C}) \cite{dft} is better, nevertheless, careful consideration of the free parameters of the theory is required. Besides, even when these models were theoretically consistent, the cosmological constant term may be needed in general for their cosmological viability ($\alpha\neq 0$). As the authors of \cite{dft} recognize, the need of such a constant term makes these models unappealing from a theoretical point of view.

\section{Conclusions}\label{conclusions}

In this paper we have performed a careful study of the stability of vacuum linearization of several cosmologically viable models based on quadratic modifications of gravity. This kind of study is complementary to other existing tests such as the application of the dynamical systems tools \cite{zcs}, and model-fitting of the combined data sets \cite{cosmo tests}. We have demonstrated that several models that pass the latter tests, fail to have a consistent vacuum linearization due to the instabilities caused by the presence of ghost and tachyon states. On the basis of the mere occurrence of any of these instabilities a given, supposedly viable cosmological model, might be rule out. 

We have paid special attention to: i) the development of ghost graviton, associated with increasing values of the effective Newton's constant $G_N^{eff}$ with the background curvature (the so called Dolgov-Kawasaki instability, also called Ricci instability), ii) to the requirement of non-negativity of the effective gravitational coupling $G_{eff}$, ensuring that the graviton is not a ghost state, and iii) to absence of a destabilizing spin-0 tachyon perturbation.

The results of our study suggest that models (\ref{zcs A}-\ref{zcs C}) \cite{zcs} exhibit vacuum instabilities of the kind considered if, at the same time, the best-fit values of the free parameters taken from \cite{cosmo tests} are considered. While models (\ref{dft A}-\ref{dft C}) are less problematic in this respect -- although a careful study of the parameter space is necessary --, the models containing powers of the GB invariant \cite{navarro,odintsov2} are clearly to be rule out. 

Our conclusion is that the inclusion of quadratic modifications of gravity, the way these modifications have been considered so far, results in very unappealing models. Perhaps an alternative simple way to introduce these modifications is through the Dirac-Born-Infeld deformation strategy, according to which the Lagrangian density ${\cal L}=\sqrt{|g|}L$ is to be replaced by ${\cal L}=\sqrt{|g|}\epsilon\lambda(\sqrt{1+2\epsilon L/\lambda}-1)$ ($\epsilon=\pm 1$, $\lambda$ is an additional energy scale). This strategy has been invoked to erase unpleasant singularities from the original theory. In the context of the quadratic theories of gravity it might help to avoid or smooth out several of the instabilities considered here \cite{quiros-urena}. This will be the subject of future publications.

This work was partly supported by CONACYT, grant number I0101/131/07 C-234/07, Instituto Avanzado de Cosmologia (IAC) collaboration. E T acknowledges the MES of Cuba by partial support of this research.

\end{document}